\documentclass[journal]{IEEEtran}

\ifCLASSINFOpdf
\else
\fi

\usepackage{graphicx}
\usepackage{epstopdf}
\usepackage{multicol}
\usepackage{url}
\usepackage{amsmath}
\usepackage{mathtools}
\usepackage{amssymb}
\usepackage{esint}
\usepackage{amsfonts}
\usepackage{multirow}
\usepackage{wrapfig}
\usepackage{hyperref}
\usepackage{cite}
\usepackage{subcaption}
\usepackage{lipsum}
\usepackage{float}	
\usepackage{caption}
\usepackage{algorithm}
\usepackage{algorithmic}
\usepackage[none]{hyphenat}
\usepackage[usenames,dvipsnames]{color}
\usepackage{wrapfig,lipsum,booktabs}

\begin{document}

\title{\LARGE Indoor UAV Navigation to a Rayleigh Fading Source Using Q-Learning\thanks{Bekir S. Ciftler (bcift001@fiu.edu) is with Florida International University, Adem Tuncer (adem.tuncer@yalova.edu.tr) is with Yalova University, and Ismail Guvenc (iguvenc@ncsu.edu) is with North Carolina State University.}}
\author{\IEEEauthorblockN{Bekir Sait \c{C}iftler, \IEEEmembership{Student Member, IEEE}, Adem Tuncer, and
\.{I}smail G\"{u}ven\c{c}, \IEEEmembership{Senior Member, IEEE}}\vspace{-0.8cm}}
\maketitle
\begin{abstract}
Unmanned aerial vehicles (UAVs) can be used to localize victims, deliver first-aid, and maintain wireless connectivity to victims and first responders during search/rescue and public safety scenarios.
In this letter, we consider the problem of navigating a UAV to a Rayleigh fading wireless signal source, e.g. the Internet-of-Things (IoT) devices such as smart watches and other wearables owned by the victim in an indoor environment. 
The source is assumed to transmit RF signals, and a Q-learning algorithm is used to navigate the UAV to the vicinity of the source.
Our results show that the time averaging window and the exploration rate for the Q-learning algorithm can be optimized for fastest navigation of the UAV to the IoT device.
As a result, Q-learning achieves the best performance with smaller convergence time overall.
\end{abstract}


\section{Introduction}

Search and rescue, public safety, and emergency management applications may require navigation of first responders to a victim's location. 
\textcolor{black}{This can be achieved by using the signals radiated from Internet of Things (IoT) devices carried by the victims such as mobile phones, smart watches,  fitness trackers, or other smart sensors, which can be probed to send RF signals repetitively}~\cite{7572034,iotpublicsafety}.
To this end, unmanned aerial vehicles (UAVs) have been recently gaining more attention due to communication, autonomous navigation, and video capture capabilities, and they can help in localizing people during emergency situations.
For example, they can be quickly deployed within a building on fire to localize victims and first responders, to deliver first aid kits, and to maintain wireless connectivity with them for enabling real-time situational awareness through live video.

\begin{figure}[t]
\centering{\includegraphics[width=0.85\linewidth]{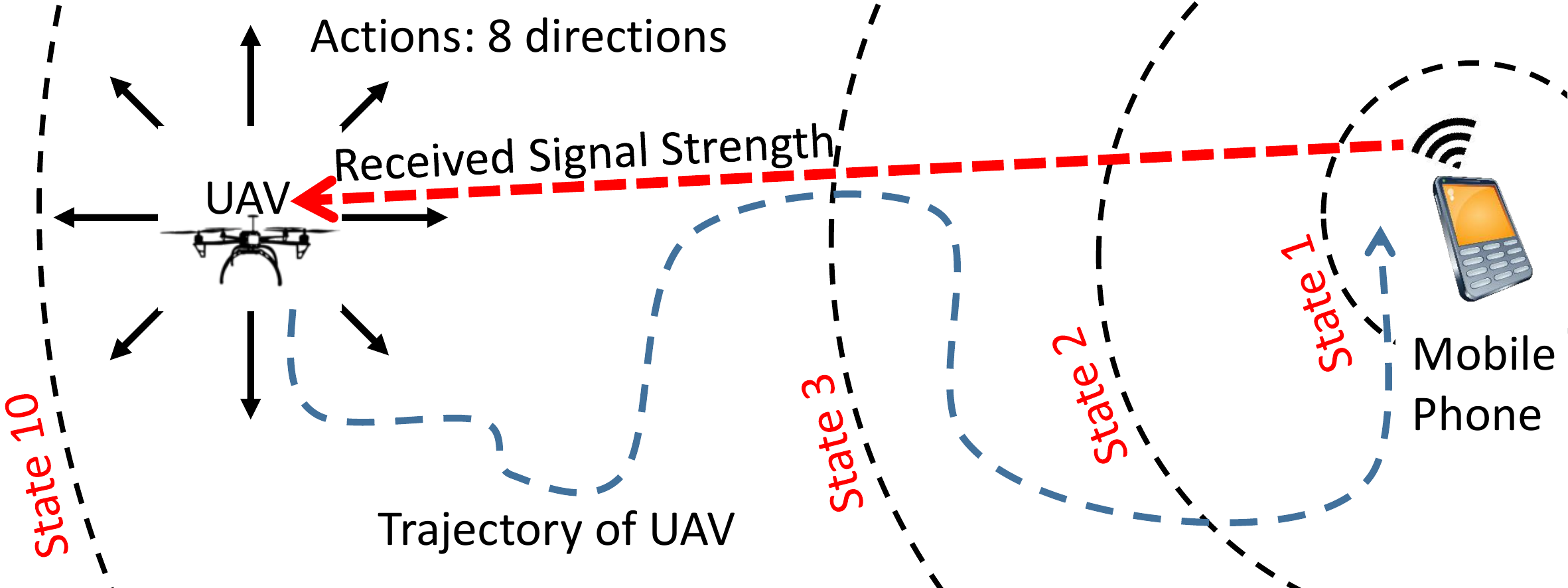}}
    \caption{Navigation to a Rayleigh fading wireless source.}
    \label{fig:System}
    \vspace{-5mm}
\end{figure}

In this letter, as shown in Fig.~\ref{fig:System}, 
we consider the problem of navigating a UAV to a Rayleigh fading RF source.
We consider a GPS-denied indoor environment and assume that the RF source continuously radiates signals. \textcolor{black}{For example, most mobile equipment continuously transmit WiFi probe requests to discover nearby access points~\cite{Vattapparamban}, and a mobile device may also be \emph{forced} by a UAV to transmit wireless signals in case of emergency incidents~\cite{2}.}
\textcolor{black}{In the literature, collaborative localization of a moving RF source is presented in~\cite{farshad} by a swarm of UAVs based on the D-optimality criteria.
Another study in~\cite{flying} presents an optimal flying path for UAV-assisted IoT sensor networks using a location aware multi-layer information map. 
It considers different utility functions based on the sensor density, energy consumption, flight time, and flying risk level, and weighted sum of multi-objective utility functions is maximized using a genetic algorithm.}
Several other works in the literature consider Q-learning and other reinforcement learning (RL) techniques for navigation of robots~\cite{Atanasov,Fink}.
To our best knowledge, there are no studies that consider the use of Q-learning for navigation of UAVs based on the received signal strength (RSS) observations from a \emph{Rayleigh fading} RF source. {\color{black}
In particular, in indoor environments, Rayleigh fading signal from the source may cause significant variations in the RSS model, and hence it can bias the navigation algorithms while deciding on the optimum actions for the UAV. On the other hand, Q-learning is a model-free reinforcement learning technique which avoids bias in the navigation of UAV.}

In this letter, we study the behavior of Q-learning based UAV navigation under Rayleigh fading assumption, and investigate averaging of the RSS over different time spans considering different UAV speeds. 
We also study a variable learning rate technique, which is shown to provide better convergence time in reaching to the Rayleigh fading source when compared with a fixed learning rate technique.
\textcolor{black}{We compare the proposed algorithm with an existing Reinforcement Learning (RL) technique~\cite{hester2013texplore}. Contributions of this work can be summarized as follows:}
\begin{itemize}
\item A model-free Q-learning algorithm for indoor UAV navigation to a Rayleigh fading RF source is proposed
\item Proposed algorithm is compared with an existing RL based technique
\item Varying and fixed learning rates are studied for convergence time
\item Various UAV speeds are studied for convergence time
\end{itemize}


\begin{figure}[t]
\centering{\includegraphics[width=0.85\linewidth]{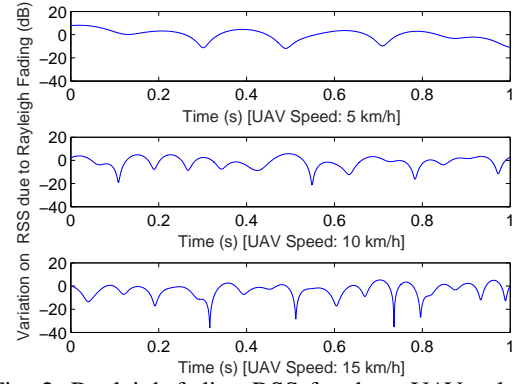}}
\vspace{-5mm}
    \caption{Rayleigh fading RSS for three UAV velocities.}
    \label{fig:Rayleigh}
    \vspace{-5mm}
\end{figure}


\smallskip 

\noindent \textit{\textbf{System Model:}} The RSS at a UAV from a Rayleigh fading wireless source  can be calculated using the distance between the source and the UAV as 
$P_{\rm R}=P_{\rm Tx}-PL(d)-S$, 
where $P_{\rm R}$ is the RSS, $P_{\rm Tx}$ is the transmit power of the source, $PL(d)$ is the path loss at distance $d$, and $S$ is a random variable that captures Rayleigh fading which may cause deep fades in the RSS.
In this letter, we consider 3GPP TR~36.814 path loss model~\cite{Access3GPP}, given by
$PL(d)=128.1+37.6\log_{10}(d/1000)$. 

As an example, in Fig.~\ref{fig:Rayleigh}, we present Rayleigh fading RSS for three different UAV velocities, considering a fixed RF source.
Results show that the RSS may observe deep fades which may cause as large as $40$~dB losses for certain cases.
Moreover, the variation of the RSS increases with larger UAV speeds as shown in Fig.~\ref{fig:Rayleigh}.
Such deep fades may bias the UAV that it may be navigating in the wrong direction, even when the UAV may be approaching closer to the target node.
In the next section, we will present the proposed Q-learning algorithm, and discuss how the Rayleigh fading effects as in Fig.~\ref{fig:Rayleigh} can be mitigated for navigation to target.

\section{Q-Learning Based Source Tracking}

We consider the use of Q-learning~\cite{Konar,Watkins} 
algorithm to navigate the UAV towards the wireless source based on the RSS observations at the UAV.
Q-learning is an improved RL technique which can operate without any prior knowledge about the environment or the model for RSS observations.
It learns by trial and error, and iteratively builds a value function of each state-action pair.
The goal is to select the action which has maximum  $Q$-value using following update rule at~each~step:
\begin{align}
Q(s,a)\leftarrow Q(s,a)+\alpha\big[r(s,a)+\gamma\max\limits_{a'}Q(s',a')-Q(s,a)\big]~,\label{eq:q_fn}\nonumber
\end{align}
where $s'$ is the state reached from state $s$ after action $a$, $\alpha \in [0,1]$ is the learning rate to control learning speed, and  $r(s,a)$ is the immediate reward received as result of action $a$.
We use two different learning rate models: a fixed learning rate with $\alpha$ set to a constant, $0.5$, and a varying learning rate where $\alpha$ is dynamically modified based on the observations. In varying learning rate model, as the quality of signal gets better (i.e. UAV is closer to source) learning rate increases not to miss or pass by the source.
The discount factor is represented with $\gamma \in [0,1]$, which determines the importance of future rewards. 

\begin{wraptable}{l}{4.1cm}\small 
\vspace{-1mm}
\centering
	{\begin{tabular}{|c|c|}
			\hline 
            {\bf State} & {\bf RSS (dBm)} \\ \hline
	$s$=1	& $P_{\rm R}>-40$\\
\hline
	$s$=2	&			$-50\leq P_{\rm R} \leq$ $-40$\\\hline
	$s$=3	&			$-60\leq P_{\rm R} \leq$ $-50$\\\hline
	$s$=4	&			$-70\leq P_{\rm R} \leq$ $-60$\\\hline
	$s$=5	&			$-80\leq P_{\rm R} \leq$ $-70$\\\hline
	$s$=6	&		    $-90\leq P_{\rm R} \leq$ $-80$\\\hline
	$s$=7	&			$-100\leq P_{\rm R} \leq$ $-90$\\\hline
	$s$=8	&			$-110\leq P_{\rm R} \leq$ $-100$\\\hline
	$s$=9	&			$-120\leq P_{\rm R} \leq$ $-110$\\\hline
	$s$=10	&			$P_{\rm R}<-120$\\\hline
\end{tabular}}
        \caption{UAV states  with respect to source RSS.}
        \label{Tab:1}\vspace{-4mm}
\end{wraptable}

Balancing of exploration and exploitation is a critical issue in RL techniques~\cite{Konar} and there are several strategies to maintain this balance, such as $\epsilon$-greedy and softmax approaches~\cite{Chen}. 
In here, the $\epsilon$-greedy exploration strategy is used, where  $\epsilon\in [0,1]$ is the exploration probability. While the action is generally selected at each step according to highest Q-value for exploitation, for exploration the selection is carried out randomly with a small probability $\epsilon$.
We consider that the RSS observed at the UAV belongs to a finite set $s\in\{1,...,10\}$ of 10 different states as in Table~\ref{Tab:1}. 

The pseudo-code of the Q-learning technique that we used is presented in Algorithm~\ref{alg:Alg1}. {\color{black} We consider two dimensional mobility and eight actions $a\in\{1,...,8\}$ for a UAV corresponding to eight uniformly-spaced directions in angular domain as shown in Fig.~\ref{fig:System}. 
This is because we assume that the UAV flies at a fixed height, and that the distance between the floor and the ceiling are insignificant compared to the distance between the UAV and the RF source.}
The Q-values of each state-action pair are then stored in a $10\times 8$ matrix, initialized to zero at the beginning of the algorithm, and populated with the value of each specific action as new observations are obtained. The reward is set as the difference between the latest two values of the RSS, which increases the likelihood of choosing actions that will move the UAV towards the target. On the other hand, due to Rayleigh fading, UAV may also occasionally choose erroneous actions.

\begin{algorithm}[t]\small 
	\caption{Q-learning based indoor UAV navigation.}
    \label{alg:Alg1}
	\begin{algorithmic}[1]
		\STATE initialize all $Q(s,a)$ table to	zero
		\STATE \textbf{repeat} (for each step):
		\STATE \hspace{0.3cm} obtain RSS and average the last three RSS
		\STATE \hspace{0.3cm} obtain $r(s,a)$ and $s$ according to averaged RSS
		\STATE \hspace{0.3cm} choose $a$ from $s$ using policy from $Q$ using $\epsilon$-greedy
		\STATE \hspace{0.3cm} take action $a$, observe $r(s,a)$~and~$s'$
		\STATE \hspace{0.3cm} check the new location using action $a$ for obstacle(s)
		\STATE \hspace{0.3cm} \textbf{while} any obstacle at new location with action $a$ \textbf{do}
		\STATE \hspace{0.6cm} leave $a$ and select any other action $a$ randomly, \textbf{end}
		\STATE \hspace{0.3cm} update $Q$-value according to equation (1)
        \STATE \hspace{0.3cm} \textbf{if} varying learning rate
		\STATE \hspace{0.6cm} update $\alpha \in [0,1]$ values according to $s'$, \textbf{end}
		\STATE \hspace{0.3cm} $s \leftarrow s'$
		\STATE \textbf{until} $s$ is terminal
	\end{algorithmic}
\end{algorithm}

In order to minimize erroneous decisions due to deep fades in the RSS (see Fig.~\ref{fig:Rayleigh}), we average the RSS over a sampling duration $T_{\rm S}$ before feeding it into the Q-learning algorithm. 
While a large averaging window $T_{\rm S}$ will help in minimizing wrong decisions due to deep fades, it will also introduce delays in choosing a new action value, and hence may delay the convergence time. On the other hand, a shorter averaging window will enable more frequent actions, albeit with more likelihood of RSS being subject to deep fades. Therefore, there is an optimum averaging window duration, which will result in the fastest navigation of the UAV to the RF source. 

\begin{figure}[t]	
	\centering{\includegraphics[width=0.85\linewidth]{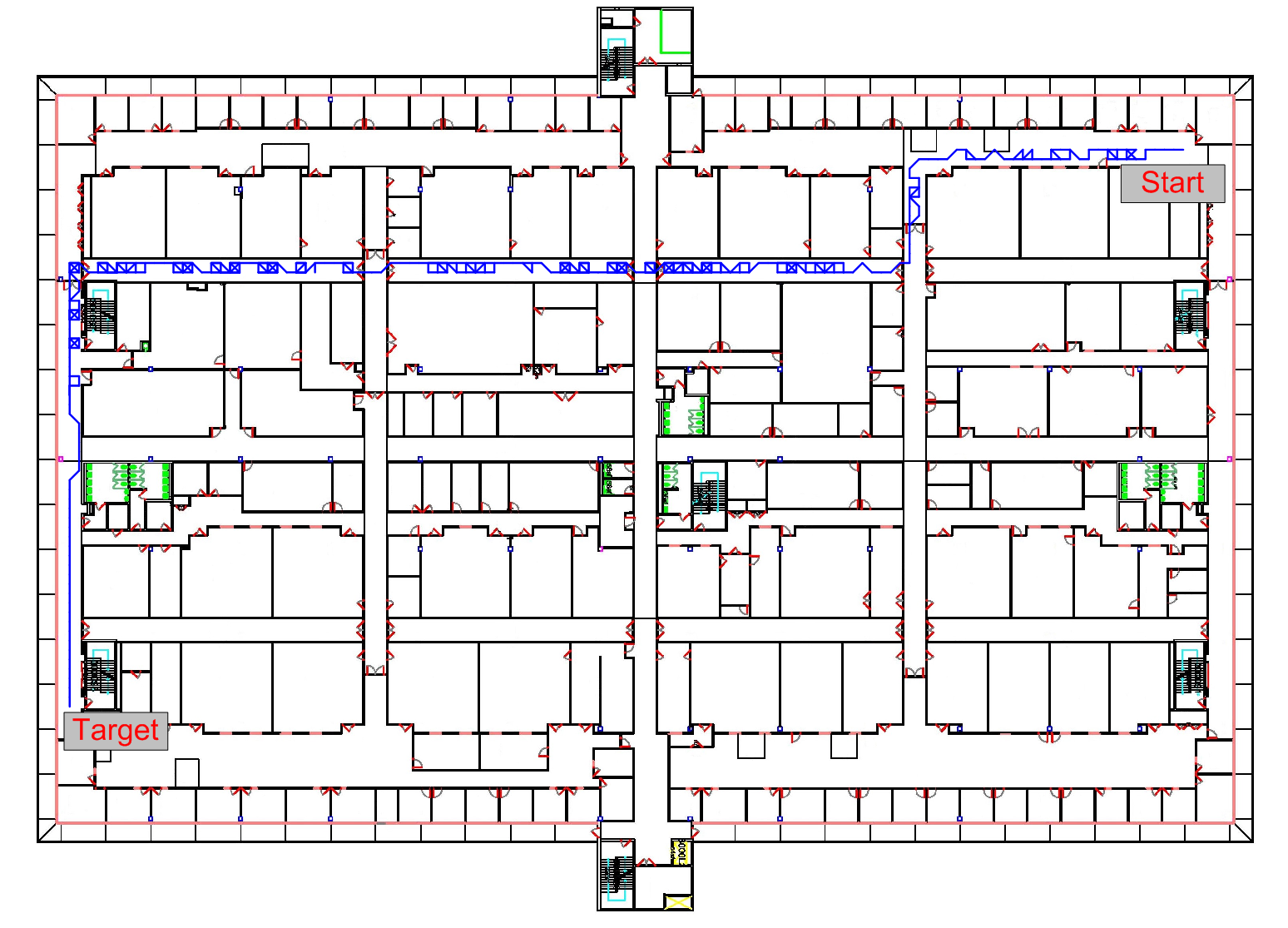}}\vspace{-1mm}
	\caption{Map of FIU Engineering Center 3rd floor.}
	\label{fig:FIUECMAP}
    \vspace{-5mm}
\end{figure}

\begin{figure*}[t]
\centering
\includegraphics[width=1.7\columnwidth]{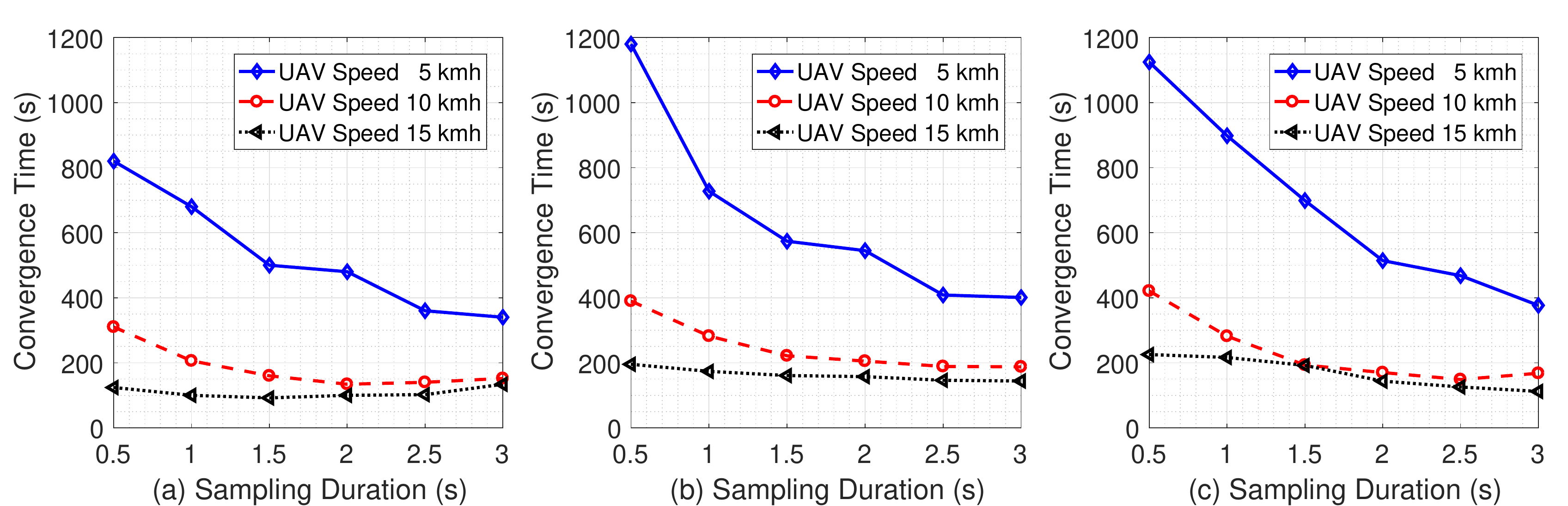}
\caption{Convergence time of (a) Q-learning (varying learning rate); (b) Q-learning (fixed learning rate); (c) RL (single state).}
\label{fig:ResultPath}
\end{figure*}


\section{Simulation Results and Conclusion} 

\textcolor{black}{The map of the FIU EC 3rd Floor ($75$~m by $120$~m) is used for simulations as shown in Fig.~\ref{fig:FIUECMAP}.}
It is assumed at the beginning of the simulation that the UAV and the RF source are positioned at the opposite corners of the floor considering a worst case scenario. The UAV can only navigate through the aisles on the map, without crashing into any walls. 
If any possible crash is detected for a selected action (e.g., through sonar sensors), the UAV tries a different action.
The RSS is assumed to vary based on Rayleigh fading caused by the velocity of UAV and path loss dictated by the distance between the UAV and the target node.

Simulation results\footnote{MATLAB codes that are used to generate the results in this paper are publicly available at \url{https://research.ece.ncsu.edu/mpact/data-management/}.} are obtained under several sampling durations, a variable (heuristically optimized) learning rate $\alpha$, and three different UAV speeds, as shown in Fig.~\ref{fig:ResultPath}.
{\color{black}In~Fig.~\ref{fig:ResultPath}(a) shows the convergence time for Q-learning with variable learning rate. The time it takes for the UAV to reach the vicinity of the target (i.e., the convergence time) tends to decrease with larger sampling duration for lower speeds. 
On the other hand, for high velocities, the convergence time starts increasing for larger sampling duration, since longer distances will be traveled by the UAV in case of wrong decisions.
In general, larger velocity, at least for the considered set of three UAV velocities in this letter, results in faster convergence to the target node. 

Use of fixed learning rate has a noticeable impact on the convergence performance of the Q-learning algorithm.
In particular, Fig.~\ref{fig:ResultPath}(b) shows results with same UAV and target locations, and we present the convergence performance when the learning rate is fixed to $1$.
Using a fixed learning rate decreases responsiveness of the system. Hence, the convergence time increases significantly.

Finally, the proposed Q-learning algorithm is compared with the RL-based technique in~\cite{hester2013texplore}. In Fig.~\ref{fig:ResultPath}(c), the results for the RL-based technique is given.
Although it achieves a similar performance to our proposed algorithm at higher velocities, our proposed algorithm performs better at lower velocities.
Since the emergency situations may require delicate actions to avoid dangerous situations for a victim or first responder, lower velocities may be preferred in most cases.

In general, our overall results prove that it is critical to use a Q-learning based approach for avoiding the navigation bias in a Rayleigh fading environment with sufficiently large window. Another observation is that a variable learning rate is preferable compared to a fixed learning rate for increasing responsiveness of the system.}\\

\textit{\textbf{Acknowledgement:} This research is supported in part by the National Science Foundation grant AST-1443999.}







\begin{thebibliography}{}


\bibitem{7572034}
N.~H. Motlagh, T.~Taleb, and O.~Arouk, ``Low-altitude unmanned aerial
  vehicles-based Internet of Things services: Comprehensive survey and future
  perspectives,'' \emph{IEEE Internet of Things Journal}, vol.~3, no.~6, pp.
  899--922, Dec 2016.

\bibitem{iotpublicsafety}
Chunquan, D.,~and~Zhu, S.: `Research on urban public safety emergency management early warning system based on technologies for the Internet of things.' \textit{Procedia Engineering }, 2012, {45}, pp. 748--754.
	
\bibitem{Vattapparamban}	
Vattapparamban,~E., Ciftler,~B.~S, Guvenc,~I., Akkaya,~K., Kadri,~A.: `Indoor Occupancy Tracking in Smart Buildings Using WiFi Probe Requests', in Proc. IEEE ICC Workshops, Malaysia, May 2016.

\bibitem{2}	Wang,~W., Joshi,~R., Kulkarni,~A., Leong,~W.~K., Leong,~B.: `Feasibility study of mobile phone WiFi detection in aerial search and rescue operations,' in Proc. Asia-Pacific Wkshp. Syst., NY, 2013, pp~1--6.

\bibitem{farshad}
Koohifar,~F., Kumbhar,~A., Guvenc,~I.: `Receding Horizon Multi-UAV Cooperative Tracking of Moving RF Source', in preprint, accepted by \textit{IEEE Communication Letters}, 2016


\bibitem{flying}
Yoo,~S., Park,~J., Kim,~S., Shrestha,~A.:`Flying path optimization in UAV-assisted IoT sensor networks', Elsevier ICT Express, 2016, vol. 2, num. 3, pp 140--144

\bibitem{Atanasov}
Atanasov,~N., Le Ny,~J., Michael,~N., Pappas,~G.~J.: `Stochastic source seeking in complex environments', Robotics and Automation (ICRA), IEEE Int. Conf. on, Saint Paul, MN, USA, May 2012, pp 3013--3018.

\bibitem{Fink}
Fink,~J., Kumar,~V.: `Online methods for radio signal mapping with mobile robots', Robotics and Automation (ICRA), IEEE Int. Conf. on, Anchorage, AK, USA, May 2010, pp 1940--1945.

\bibitem{hester2013texplore}
T.~Hester and P.~Stone, ``Texplore: real-time sample-efficient reinforcement
  learning for robots,'' Springer, \emph{Machine learning}, vol.~90, no.~3, pp. 385--429,
  2013.

\bibitem{Access3GPP}
Access, Evolved Universal Terrestrial Radio. `Further advancements for E-UTRA physical layer aspects', vol. 9. 3GPP TR 36.814, 2010.

\bibitem{Watkins}
Watkins, C., Dayan, P.: `Q-Learning', \textit{Machine Learning}, 1992, {8}, pp. 279--292

\bibitem{Konar}
Konar,~A., Chakraborty,~I.~G., Singh,~S.~J., Jain,~L.~C., Nagar,~A.~K.: `A Deterministic Improved Q-Learning for Path Planning of a Mobile Robot', \textit{IEEE Trans. Syst., Man, Cybern., Syst.}, 2013, {43}, (5), pp. 1141--1153.

\bibitem{Chen}
Chen,~C., Dong,~D., Li,~H.~X., Chu,~J., Tarn,~T.~J.: `Fidelity-Based Probabilistic Q-Learning for Control of Quantum Systems', \textit{IEEE Trans. Neural Netw. Learn. Syst.}, 2014, {25}, (5), pp. 920--933.


\end{thebibliography}
\end{document}